\documentclass{moriond}

\usepackage{bm}
\usepackage{amsmath}
\usepackage{amssymb}
\usepackage{comment}

\def\Journal#1#2#3#4{{#1} {\bf #2}, #3 (#4)}

\def\PRL{\em Phys. Rev. Lett.}
\def\PRD{{\em Phys. Rev.} D}

\def\METRO{\em Metrologia}
\def\SCI{\em Science}
\def\NAT{\em Nature}
\def\NATCOM{\em Nat. Commun.}
\def\NJP{\em New J. Phys}

\def\PRA{{\em Phys. Rev.} A}

\def\CQG{\em Class. Quantum Grav.}
\def\RPP{\em Reports on Progress in Physics}
\def\JOA{{\em J. Opt.} A}
\def\APB{{\em Appl. Phys.} B}

\def\be{\begin{equation}}
\def\ee{\end{equation}}
\def\bea{\begin{eqnarray}}
\def\eea{\end{eqnarray}}

\def\Eq{Eq.~}

\def\Fig{Fig.~}

\def\ie{\textit{i.e.}~}

\newcommand{\ket}[1]{\left| #1 \right\rangle}

\newcommand{\Photo}{}

\begin{document}
\vspace*{4cm}

\title{Studies of general relativity with quantum sensors}

\author{G. Lef\`{e}vre$^{1,6}$, G. Condon$^{1}$, I. Riou$^{1,6}$, L. Chichet$^{1}$, M. Essayeh$^{1,6}$, M. Rabault$^{1}$, L. Antoni-Micollier$^{1}$,\\
N. Mielec$^{2,6}$, D. Holleville$^{2,6}$, L. Amand$^{2,6}$, R. Geiger$^{2,6}$, A. Landragin$^{2,6}$, M. Prevedelli$^3$,\\
B. Barrett$^{1,4}$, B. Battelier$^{1}$, A. Bertoldi$^{1,6}$, B. Canuel$^{1,5,6}$, and P. Bouyer$^{1,6}$}

\address{$^1$LP2N, Laboratoire Photonique, Num\'{e}rique et Nanosciences, Universit\'{e} Bordeaux--IOGS--CNRS:UMR 5298, rue F. Mitterrand, F--33400 Talence, France}

\address{$^2$LNE--SYRTE, Observatoire de Paris, PSL Research University, CNRS, Sorbonne Universit\'{e}s, UPMC Univ. Paris 06, 61 avenue de l'Observatoire, F--75014 Paris, France}

\address{$^3$Dipartimento di Fisica e Astronomia, Universit{\`a} di Bologna, Via Berti-Pichat 6/2,\\
I--40126 Bologna, Italy}

\address{$^4$iXblue, 34 rue de la Croix de Fer, 78100 Saint-Germain-en-Laye, France}

\address{$^5$LSBB, Laboratoire Souterrain {\`a} Bas Bruit UNS, UAPV, CNRS:UMS 3538, AMU,\\
La Grande Combe, F--84400 Rustrel, France}

\address{$^6$MIGA Consortium}

\maketitle

\abstracts{We present two projects aiming to probe key aspects of the theory of General Relativity with high-precision quantum sensors. These projects use cold-atom interferometry with the aim of measuring gravitational waves and testing the equivalence principle. To detect gravitational waves, a large multi-sensor demonstrator is currently under construction that will exploit correlations between three atom interferometers spread along a 200 m optical cavity. Similarly, a test of the weak equivalence principle is currently underway using a compact and mobile dual-species interferometer, which will serve as a prototype for future high-precision tests onboard an orbiting satellite. We present recent results and improvements related to both projects.}


\section{Introduction}

Over the past few decades, the development of laser-cooling techniques~\cite{Chu1997} has made it possible to exploit the quantum properties of matter at very low temperatures. These techniques have enabled experimentalists to coherently manipulate quantum objects with a very high degree of precision. In this context, atom interferometry has emerged as a powerful tool for metrology. Nowadays, atom interferometers (AIs) are used for a wide range of applications, such as sensitive probes of inertial forces~\cite{metro1,prl1,prl2,prl3}, or studies of fundamental physics \cite{sci1,Rosi2014,prl4,Burrage2015} and tests of gravitational theories~\cite{Barrett2015,prl5}.

In this frame, we are developing two AI experiments that exploit differential measurement techniques to study two aspects of the theory of General Relativity (GR). The MIGA (Matter-wave laser Interferometer Gravitation Antenna) project~\cite{spie1} aims to build a demonstrator for future large-scale gravitational wave (GW) detectors based on atom interferometry. The instrument will consist of a set of three AIs (each using a cold source of $^{87}$Rb) that are simultaneously interrogated by the resonant field of a 200 m cavity. The ICE (Interf\'erom\'etrie atomique \`a sources Coh\'erentes pour l'Espace) project~\cite{Barrett2015} uses a dual-species gravimeter to test the weak equivalence principle (WEP). Here, two overlapped samples of $^{39}$K and $^{87}$Rb are simultaneously interrogated during free-fall---yielding a precise measurement of their differential acceleration under gravity. These experiments have been carried out on ground, and most recently in the weightless environment of parabolic flight~\cite{Barrett2016} with the aim of increasing the measurement sensitivity, as well as serving as a test bed for future experiments in Space.


\section{Principles of the Mach-Zehnder atom interferometer}
\label{section:MachZehnder}

In both the MIGA and ICE projects, the AIs utilize a sequence of three light pulses to coherently split, reflect and recombine the atoms. Using two different atomic states, $\ket{\alpha}$ representing the ground state and $\ket{\beta}$ representing the excited state with an extra momentum transferred by the light, this sequence of light-matter interactions force the atoms to follow two separate trajectories that form a closed path as in an optical Mach-Zehnder interferometer. The sensitivity of the AI to inertial effects scales as the space-time area enclosed by the atoms. The Mach-Zehnder AI pulse sequence operates as follows. Atoms initially in the state $\ket{\alpha}$ are placed in a coherent superposition of states $\ket{\alpha}$ and $\ket{\beta}$ by a ``beam splitter'' pulse corresponding to an area of $\pi/2$. After a time $T$ of free evolution, the atomic population in these states are reversed are by a ``mirror'' pulse corresponding to an area of $\pi$. After the mirror pulse, the atoms experience a second free evolution time $T$ followed by a recombination stage via a final $\pi/2$-pulse. By measuring the populations of the states $N_{\alpha}$ and $N_{\beta}$ at the output of the interferometer, one can deduce the total transition probability $\ket{\alpha} \to \ket{\beta}$ given by $P = N_{\beta}/(N_{\alpha} + N_{\beta})$. This probability is related to the total atomic phase $\Delta\Phi_{\rm AT}$ accumulated by the matter waves over the two interferometer pathways
\be
  \label{phi0}
  P = \frac{1}{2}\big(1 - \mathcal{C} \cos(\Delta\Phi_{\rm AT}) \big),
\ee
where $\mathcal{C}$ is the contrast of the interference fringes. This atomic phase depends on the phase $\phi(\bm{r}_i, t_i)$ of the interrogation laser pulse $i$ that is imprinted on the matter waves at position $\bm{r}_i$ and time $t_i = \{0, T, 2T\}$
\be
  \label{philaser}
  \Delta\Phi_{\rm AT} = \phi(\bm{r}_1, 0) - 2\phi(\bm{r}_2, T) + \phi(\bm{r}_3, 2T).
\ee
For example, in the case of an atom accelerating under gravity, the atomic phase shift can be shown to be $\Delta\Phi_{\rm AT} = \bm{k}\cdot\bm{g} T^2$, where $\bm{k}$ is the effective wavevector of the interrogation light and $\bm{g}$ is the gravitational acceleration vector. The high-sensitivity of atom interferometers to inertial effects becomes clear from this expression, since the phase scales quadratically with $T$ and $k$ is large for optical transitions ($\sim 10^7$ rad/m). The AI enables one to sense any effect that modifies this phase, such as the acceleration arising from an inertial force or a phase modulation of the interrogation laser induced by a GW.

On the ICE experiment, the interrogation lasers are orientated along the vertical direction to maximize the sensitivity to gravity and allowing a comparison of the gravitational acceleration for two different atomic species. In comparison, the MIGA apparatus will use interrogation beams oriented along the horizontal in a ``gradiometer'' configuration---enabling the detection of GWs via the phase variation between two distant $^{87}$Rb AIs.


\section{The MIGA project: toward a new generation of gravitational wave detectors}

\begin{figure}[!t]
  \centering
  \includegraphics[width=0.6\linewidth]{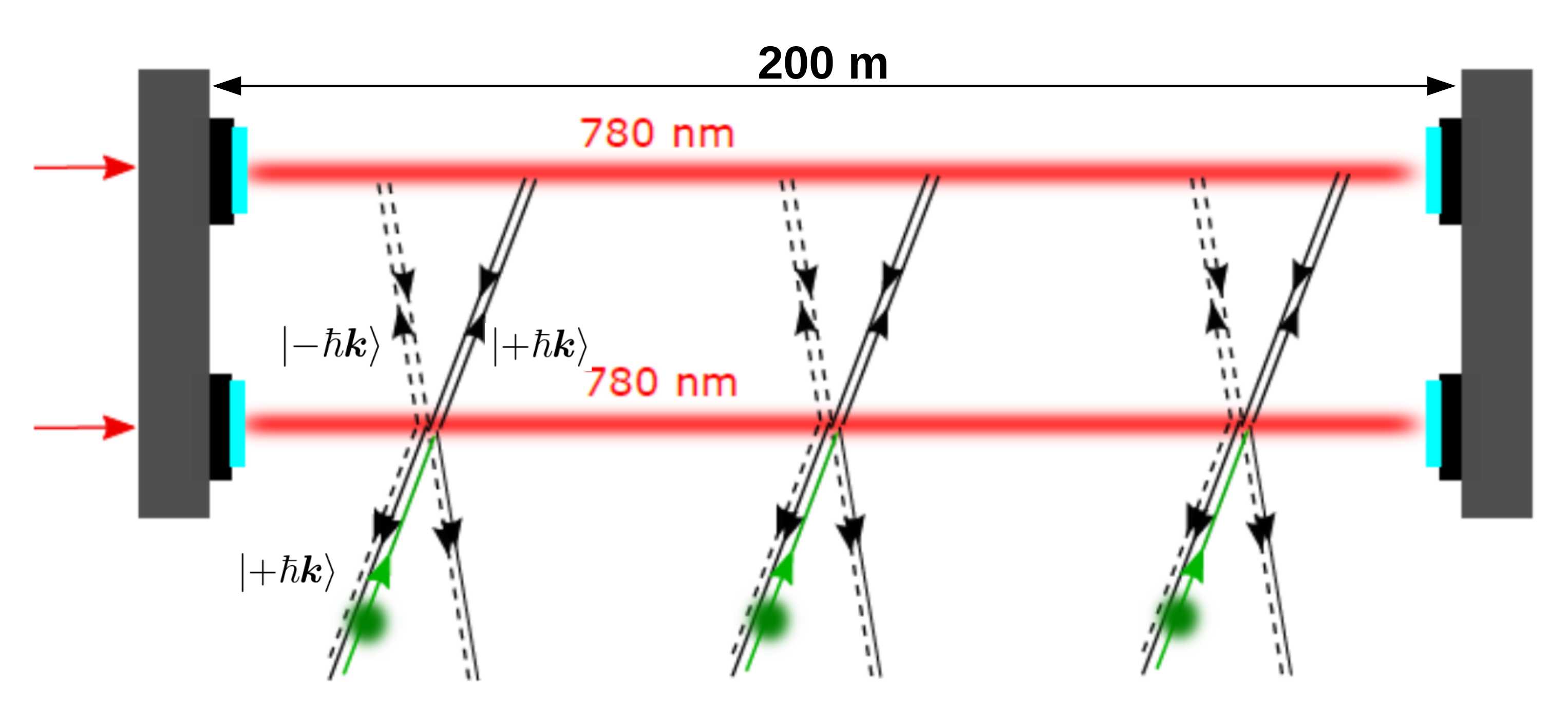}
  \caption{Scheme of the MIGA demonstrator. Atoms are vertically launched and interrogated by three Bragg pulses resonating inside two cavities. Atoms in the external state $\ket{+\hbar \bm{k}}$ first pass through the lower cavity where they are diffracted into a coherent superposition of states $\ket{+\hbar \bm{k}}$ and $\ket{-\hbar \bm{k}}$ by a $\pi/2$-pulse. At the top of the atomic trajectory, the states are reflected by a $\pi$-pulse in the upper cavity. When the atoms pass again through the lower cavity, the states are recombined by a final $\pi/2$-pulse.}
  \label{fig:GeometryOfMiga}
\end{figure}

Gravitational waves were observed for the first time in September 2015 by the LIGO detector~\cite{prl6}, which measured the transient event of a merging black hole binary system. This discovery has opened a new window to observe the universe, permitting for the first time the study of objects and events in the GW domain. This gravitational astronomy will benefit from the development of new generations of detectors that will extend the frequency band of GW detection. State-of-the-art detectors based on optical interferometry like LIGO~\cite{rpp1} or VIRGO~\cite{joa1} have a detection band starting at tens of Hz because of different sources of cavity-length noise at low frequencies. To overcome this limitation, a promising solution could rely on using correlated measurements between distant free-falling AIs. In such a configuration, the differential AI measurement enables one to read the GW phase variation induced on a common interrogation laser. The MIGA project consists of building a proof-of-principle detector by interrogating three distant AIs within the resonant field of an optical cavity. The geometry of the antenna is represented in \Fig \ref{fig:GeometryOfMiga}: three AIs in a fountain configuration are simultaneously created by three cavity-enhanced light pulses ($\pi/2 - \pi - \pi/2$ Mach-Zehnder sequence) using Bragg diffraction~\cite{prl8,pra1} in the stationary cavity field. The transitions involve the external states $\ket{+\hbar \bm{k}}$ and $\ket{-\hbar \bm{k}}$ of the $^{87}$Rb atoms in the internal ground state $\ket{F = 2, m_F = 0}$.

By measuring the atomic phase shift $\Delta \Phi_{\rm AT}(X_{i})$ from the AIs placed at different locations $X_{i}$, we can extract the strain variation $h$ induced by a GW with a strong immunity against the position noise of the cavity mirrors~\cite{ax2} using the differential measurement:
\be
  \label{GradioMeas}
  \Delta \Phi_{\rm AT}(X_{i}) - \Delta \Phi_{\rm AT}(X_{j}) \propto \frac{4\pi}{c} \frac{\nu_{0}}{2}s_{h}(X_{i} - X_{j}),
\ee
where $\nu_{0}$ is the laser frequency and $c$ is the speed of light. The term $s_{h}$ is related to weighting of time-fluctuations of $h$ by the AI sensitivity function. Assuming that (\textit{i}) the measurement of each AI is limited by quantum projection noise at a level of 1 mrad, (\textit{ii}) the maximum distance between AIs is given by the cavity length $L = 200$ m, and (\textit{iii}) the total interferometer time is $2T = 500$ ms, we obtain from \Eq \eqref{GradioMeas} a peak strain sensitivity of $2 \times 10^{-13}/\sqrt{\rm Hz}$ at a frequency of 2 Hz. To improve the measurement sensitivity, high-order Bragg diffraction~\cite{prl9} could be used. Instead of using transitions between the states $\ket{+\hbar \bm{k}}$ and $\ket{-\hbar \bm{k}}$ it is possible to induce transitions between the states $\ket{+n\hbar \bm{k}}$ and $\ket{-n\hbar \bm{k}}$ using matter-wave diffraction of order $n$. Such a process requires large optical powers that can be reached using cavity build-up.


\section{Cavity-enhanced atom interferometry}

We are currently working on a prototype to study cavity-enhanced atom interferometry in the MIGA configuration. This experiment consists of a single AI where atoms are manipulated by three Bragg pulses resonating inside two 80-cm-long cavities.


\subsection{The atomic source and interferometer}

\begin{figure}[!th]
  \centering
  \includegraphics[width=1\linewidth]{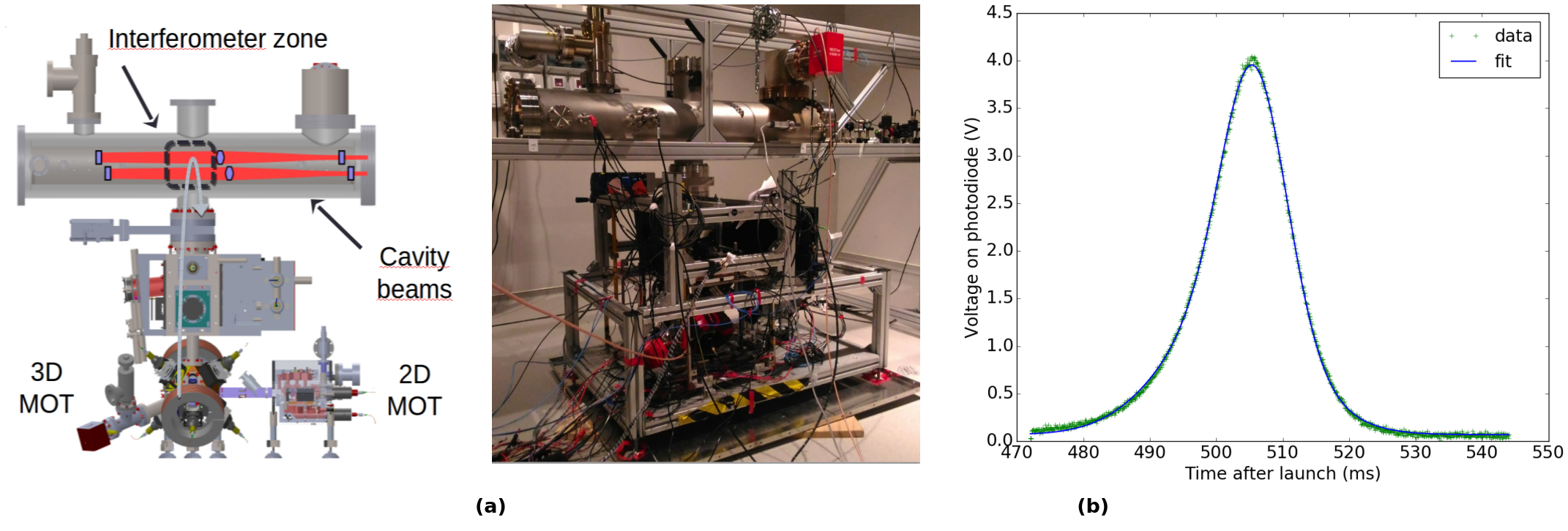}
  \caption{(a) Left: Scheme of the atom interferometer. Atoms are loaded in a 3D MOT from a 2D MOT and are launched on vertical trajectory into the interferometer zone where they are interrogated by Bragg pulses in two optical cavities. The atoms are finally detected in the region between the 3D MOT and the cavities. Right: photo of the experiment. (b) Typical time-of-flight signal measured after 505 ms by fluorescence of the atoms. By integrating the signal we deduce an atom number of $3.6 \times 10^{8}$, and the full width at half maximum of the signal gives a temperature of $4.7$ $\mu$K.}
  \label{fig:AtomSource}
\end{figure}

Atoms are first cooled down in a three dimensional magneto-optical trap (3D MOT) to a temperature of $\sim 140$ $\mu$K. To have a fast loading rate in the 3D MOT, atoms are first trapped in a 2D MOT and pushed by a laser beam to generate an atomic flux of several $10^9$ atoms/s toward the center of the 3D MOT. Atoms are launched from the 3D MOT on a vertical trajectory [see \Fig \ref{fig:AtomSource}(a)] with a velocity of 3.0 m/s by a moving molasses technique using a frequency shift between the upper and lower cooling beams. A stage of sub-Doppler cooling reduces the sample temperature to a few $\mu$K by increasing the detuning of the cooling beams from the $F = 2$ to $F'=3$ cycling transition to about $-20\,\Gamma$, where $\Gamma \simeq 6$ MHz is the transition linewidth. Figure \ref{fig:AtomSource}(b) shows a typical signal of cold atoms measured by resonant fluorescence after a time of flight of 505 ms. A temperature of $4.7$ $\mu$K is reached without shielding the source from external magnetic fields.

For the interferometer sequence, atoms need to be prepared in a well-defined velocity class and internal state. Velocity selection is performed with a horizontally-oriented  Raman pulse before the atoms reach the interferometer zone. To minimize sensitivity to ambient magnetic fields, a second Raman pulse selects the atoms in the magnetically-insensitive $m_F = 0$ state by splitting the Zeeman sub-levels with a homogeneous magnetic bias field. After this preparation phase, the atoms enter the cavities where they are interrogated using a $\pi/2-\pi-\pi/2$ pulse sequence. The $\pi$-pulse is realized at the apex of the parabolic trajectory. On their way down, the atoms pass through the detection system, where the interferometer phase shift is measured. The populations of the two states $\ket{+\hbar k}$ and $\ket{-\hbar k}$ are detected by fluorescence after a stage of state labeling is implemented using the Raman beams.


\subsection{Geometry of the interrogation cavity}

\begin{figure}[!t]
  \centering
  \includegraphics[width=1\linewidth]{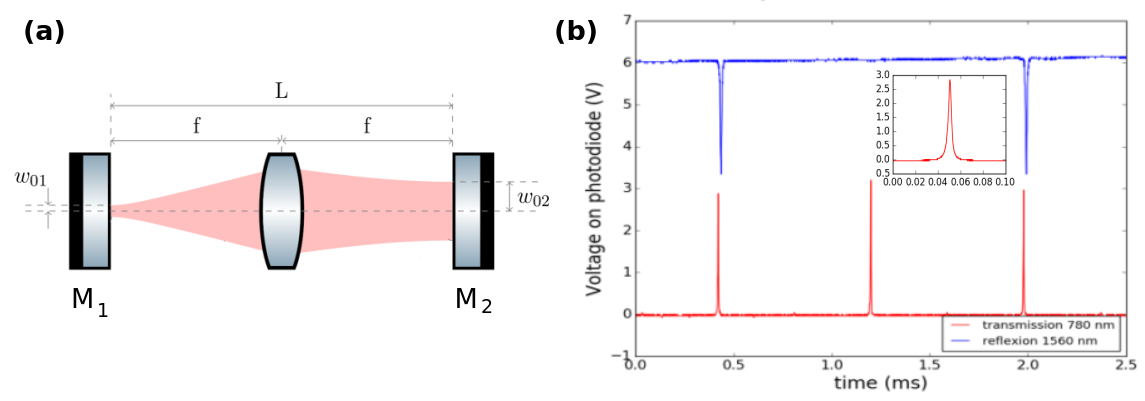}
  \caption{(a) Scheme of the marginally-stable cavity used to interrogate the atoms. Two plane mirrors with high-reflectivity coatings at 780 nm and 1560 nm are placed at the focal distance $f = 40$ cm of a biconvex lens. An input beam with waist $\omega_{01}$ will generate a beam with a magnified waist $\omega_{02}$ on the other side of the cavity where the atoms will be interrogated. (b) Cavity transmission signal at 780 nm (red line), and reflection signal at 1560 nm (blue line) obtained by scanning the current of the master laser diode operating at 1560 nm.}
  \label{fig:MargCavity}
\end{figure}

The waist of the Bragg beam needs to be of the order of several mm to interrogate efficiently the cold-atom clouds and obtain a high interference contrast. To achieve such a large waist in a stable 80-cm-long cavity, we opted for a marginally stable resonator geometry~\cite{ax1} with two plan mirrors M$_1$ and M$_2$ located at the focal point of a biconvex lens [see \Fig \ref{fig:MargCavity}(a)], which magnifies the beam size at the cavity input. If the cavity is injected with a Gaussian beam of waist $\omega_{01}$ located on M$_1$, the field resonating inside the cavity after the lens will be Gaussian, with an amplified waist $\omega_{02}$ located on M$_2$. The atom interrogation will be performed between the lens and M$_2$. The relation between the two waists is given by $\omega_{02} = \lambda f/\pi \omega_{01}$ where $\lambda$ is the wavelength of the light resonating inside the cavity.

The beam at 780 nm is used to realize the interrogation pulses of the interferometer and therefore needs to be switched off most of the time, which can lead to technical issues as it also has to be kept resonant with the cavity. The 780 nm light is therefore obtained by coherently doubling the frequency of a 1560 nm laser that is constantly locked to the cavity using the Pound-Drever-Hall (PDH) technique. However, these two beams do not resonate at exactly the same frequency inside the cavity because the mirror coating has a different effective thickness at 1560 nm and 780 nm (\ie the resonator lengths are not the same). To circumvent this problem, the 1560 nm beam is frequency modulated with an electro-optic modulator (EOM), and the PDH lock is realized on one of the modulation sidebands of this light. A resonant mode for both the sideband at 1560 nm and the carrier-doubled 780 nm light is obtained when the modulation frequency is 88.8 MHz.

\begin{figure}[!t]
  \centering
  \includegraphics[width=1\linewidth]{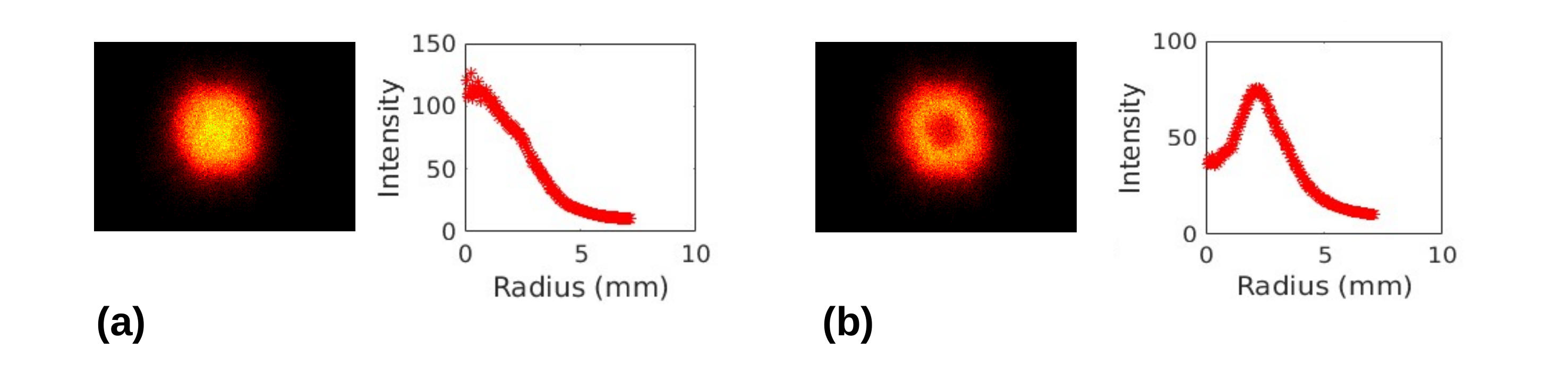}
  \caption{(a) Left: resonating mode at 780 nm (observed via cavity transmission with a CCD camera) when the cavity is perfectly aligned. Right: azimuthal average of the image, which is well described by a Gaussian intensity profile. (b) Left: resonating mode at 780 nm when the input mirror is longitudinally displaced by 50 $\mu$m. Right: azimuthal average which shows a ring-shaped intensity profile.}
  \label{fig:MargMode}
\end{figure}

Figure \ref{fig:MargCavity}(b) shows the cavity transmission at 780 nm and the reflection signal at 1560 nm used for the PHD lock, when the frequency of the 1560 nm laser is scanned by modulating its driving current. The ratio between the full width at half maximum of the resonance and the free spectral range (distance between two resonances) gives the finesse of the resonator, $F$. We obtain for the two wavelengths $F_{780} = 236$ and $F_{1560} = 180$. The height of the resonances at 780 nm gives a measure of the power transmitted by the cavity, which is the power resonating inside the cavity attenuated by the transmission of the output mirror $T_m = 0.005$. The ratio between the power inside the cavity and the input power gives the optical gain: $G_{780} = 70$. The limitation of the optical gain, which is the key parameter for future implementation of high-order Bragg diffraction, is mainly due to light diffusion on the mirrors and optical losses from the lens. A well-aligned cavity shows a Gaussian resonating mode at 780 nm, as shown in \Fig \ref{fig:MargMode}(a) from an image taken of the cavity transmission. By fitting the radial profile of the Gaussian mode, we measure a waist of $\omega_{02} = 4$ mm. As previously studied theoretically~\cite{ax1}, misalignments of the cavity optics can create an intensity variation in the beam's radial profile. For instance, for a displacement of the input mirror from the focal point of 50 $\mu$m, we observed a ring-shaped intensity distribution of the cavity field [see \Fig \ref{fig:MargMode}(b)].

With this experiment, we will soon be able to study the MIGA geometry on a reduced scale, which we anticipate will have an important impact on metrology and high-precision sensing with cavity-enhanced atom interferometry. This demonstrator will also help to implement future upgrades to the MIGA antenna, such as large-momentum-transfer atom optics.


\section{The ICE project: testing the weak equivalence principle in weightlessness}

Einstein's equivalence principle (EEP), which is a cornerstone of GR, postulates that the inertial and gravitational mass of any object are equal. This implies that, in the same gravitational field, two bodies of different masses or compositions will undergo the same acceleration. This sub-principle of the EEP is known as the weak equivalence principle (WEP), or as the universality of free fall. Various theories of quantum gravity predict a violation of the EEP~\cite{Will2006,Damour2002}, hence detecting the presence or absence of a violation at a very high precision would help to put bounds on these theories. The central parameter that characterizes the WEP is called the E\"{o}tv\"{o}s parameter $\eta$, which is defined as the ratio of the differential acceleration between two bodies to their mean acceleration
\be
  \label{eta1}
  \eta = 2\frac{|a_1-a_2|}{|a_1+a_2|},
\ee
where $a_1$ and $a_2$ are the gravitational accelerations of test bodies 1 and 2, respectively. This parameter evaluates to zero if the WEP is respected.

Today, the most precise tests of the WEP have been carried out with ``classical'' test masses (\ie masses made of bulk material) at the level of $10^{-13}$ via torsion pendulum~\cite{Schlamminger2008} or lunar laser ranging~\cite{Williams2004} experiments. Recently, the MICROSCOPE space mission~\cite{Touboul2012} has been launched and is anticipated to measure $\eta$ at $10^{-15}$ by benefiting from the low background noise and permanent free fall of an orbiting satellite. In contrast, the majority of WEP tests using cold atoms have been carried out on ground, in well-controlled laboratory environments, and have not yet reached a level of precision competitive with those done with classical bodies. Nevertheless, tests using ``quantum'' bodies like atoms are sensitive to WEP violations resulting from quantum physics that cannot be otherwise accessed. Presently, the state-of-the-art for this type of measurement has reached a precision of $3 \times 10^{-8}$ using cold samples of $^{85}$Rb and $^{87}$Rb~\cite{prl5}, or more recently to low $10^{-9}$ with $^{87}$Rb in mixtures of different internal states~\cite{Rosi2017}. So far, two main approaches have been taken to reach higher sensitivities: (\textit{i}) atoms are launched in a fountain to extend their free-fall time inside large-scale vacuum systems~\cite{Dickerson2013,Kovachy2015}, or (\textit{ii}) atoms are contained within a small-scale apparatus that is then placed in free fall, such as in a drop tower~\cite{Muntinga2013}, a sounding rocket~\cite{Seidel2012,Dinkelaker2017} or an aircraft undergoing parabolic flight~\cite{Geiger2011}.

The primary goal of the ICE project is to develop a mobile dual-species inertial sensor able to measure $\eta$ at high precision in a weightless environment, such as that generated onboard the Novespace Zero-g aircraft. Performing WEP tests in this type of environment serves as a proof-of-concept toward using cold-atom technology onboard a satellite, as proposed by the STE-QUEST mission project~\cite{Aguilera2014}, where future tests at the $10^{-15}$ level could become a reality. In the remainder of this section we will give an overview of the experiment, discuss progress toward a precision measurement on ground, and review out results from the first WEP test in weightlessness using cold-atom sensors in a free falling vehicle~\cite{Barrett2016}.


\subsection{A transportable, simultaneous dual-species interferometer}

\begin{figure}[!t]
  \centering
  \includegraphics[width=0.95\textwidth]{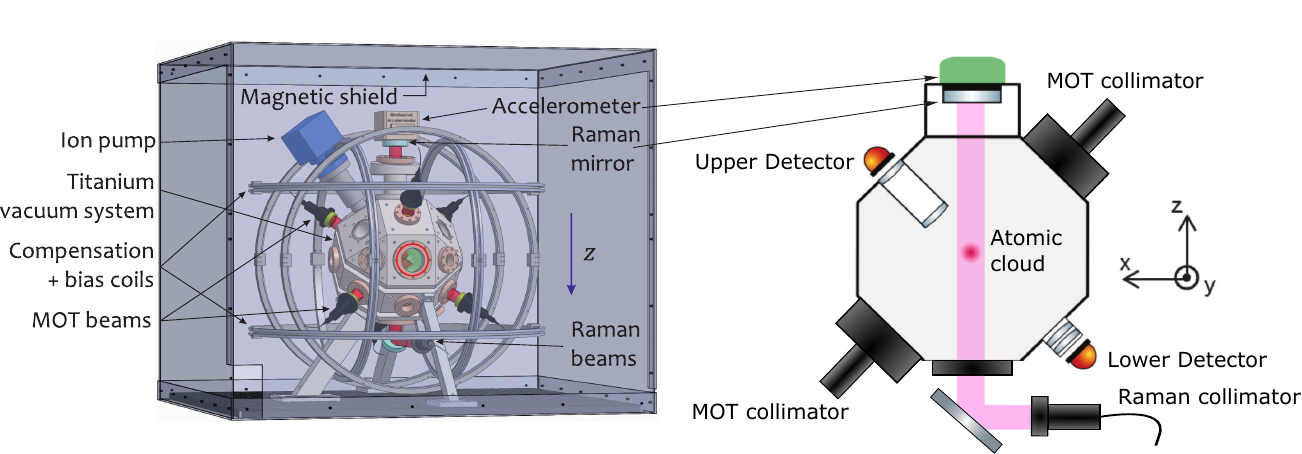}
  \caption{Scheme of the apparatus. Left: 3D view of the titanium vacuum chamber. Right: lateral section of the chamber.}
  \label{fig:vacuumchamber}
\end{figure}

The ICE project relies upon a robust and transportable apparatus comprised of a bank of fiber-based lasers, an ultra-stable frequency source and a titanium science chamber, as depicted in \Fig \ref{fig:vacuumchamber}. The potassium-39 and rubidium-87 atomic sources are derived from 3DMOTs loaded from a background vapor. After laser-cooling each sample to a temperature of a few $\mu$K and preparing the atoms in the magnetically insensitive state $\ket{F = 1, m_F = 0}$, a $\pi/2-\pi-\pi/2$ sequence of Raman pulses is applied along the vertical direction to create two Mach-Zehnder interferometers simultaneously for each species. The transition probability of each interferometer is measured via resonant fluorescence detection on the $\ket{F = 2} \to \ket{F' = 3}$ transition. This can be performed at two spatial locations by avalanche photodiodes positioned above and below the center of the chamber, as shown in \Fig \ref{fig:vacuumchamber}. The upper detector collects photons with a large solid angle using a re-entrant viewport, which is useful during $0g$ operation when the atoms do not fall, or for short interrogation times during $1g$ when the samples remain close to the chamber center. The lower detector receives a strong signal after a fixed free-fall time (up to $\sim 60$ ms), allowing for operation with larger interrogation times, and hence increased sensitivity, on ground. Finally, the interference pattern for each species is obtained by measuring the proportion of atoms in $\ket{F = 2}$ as a function of an applied laser phase that scans the sinusoidal fringe.

In a constant gravitational field, the atomic phase shift for species $j$ can be shown to be
\be
  \label{phi1}
  \Delta\Phi_j = \bm{\mathcal{S}}_j \cdot \bm{a}_j + \phi_{\mathrm{vib},j} + \phi_{\rm las},
\ee
where $\bm{\mathcal{S}}_j = \bm{k}_j T_j^2$ is the interferometer scale factor, $\phi_{\mathrm{vib},j}$ is random phase shift caused by vibrations of the retro-reflection mirror, and $\phi_{\rm las}$ is a controlled phase offset from the Raman lasers. In the absence of vibration noise, since the phase shift for each species is sensitive to the gravitational acceleration $\bm{a}_j$, one can obtain the E\"{o}tv\"{o}s parameter from
\be
  \label{eta}
  \eta = \frac{\Delta\Phi_d}{\bm{\mathcal{S}}_{\rm Rb} \cdot \bm{g}},
\ee
where $\Delta\Phi_d = |\Delta\Phi_{\rm Rb} - (\mathcal{S}_{\rm Rb}/\mathcal{S}_{\rm K}) \Delta\Phi_{\rm K}|$ is the differential phase between AIs. Equation \eqref{eta} clearly states the advantage of an interferometric WEP test---the sensitivity scales quadratically with $T$ and linearly with $k$, which can be further increased by using large momentum transfer processes~\cite{Kovachy2015}.

\begin{figure*}[!t]
  \centering
  \includegraphics[width=0.95\textwidth]{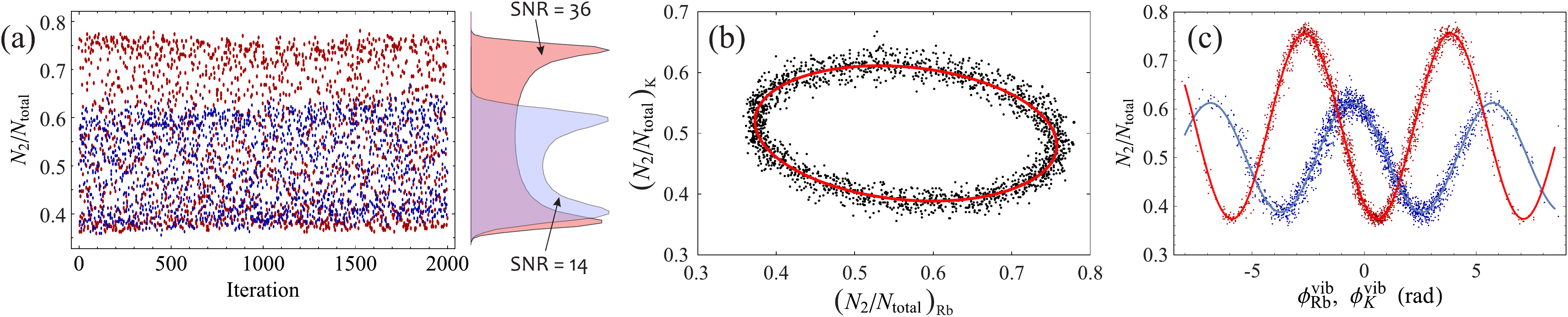}
  \caption{(a) Raw transition probability for simultaneous $^{87}$Rb (red points) and $^{39}$K (blue points) interferometers with a total interrogation time of $2T = 40$ ms. The probability density measured from these 2000 measurements is shown on the right, which enables one to extract of the fringe contrast and SNR for each species. (b) Parametric plot of the raw measurements from (a). The high degree of correlation between these data results in an ellipse. A fit to these data (solid line) yields this differential phase shift between interference fringes, which is primarily caused by systematic effects. (c) Reconstruction of the fringes using the FRAC method. Here, in additional to giving the differential phase, sinusoidal fits (solid lines) can be used to directly extract the acceleration sensitivity of each species.}
  \label{fig:KRb-CorrelatedFringes}
\end{figure*}

\begin{figure*}[!t]
  \centering
  \includegraphics[width=0.95\textwidth]{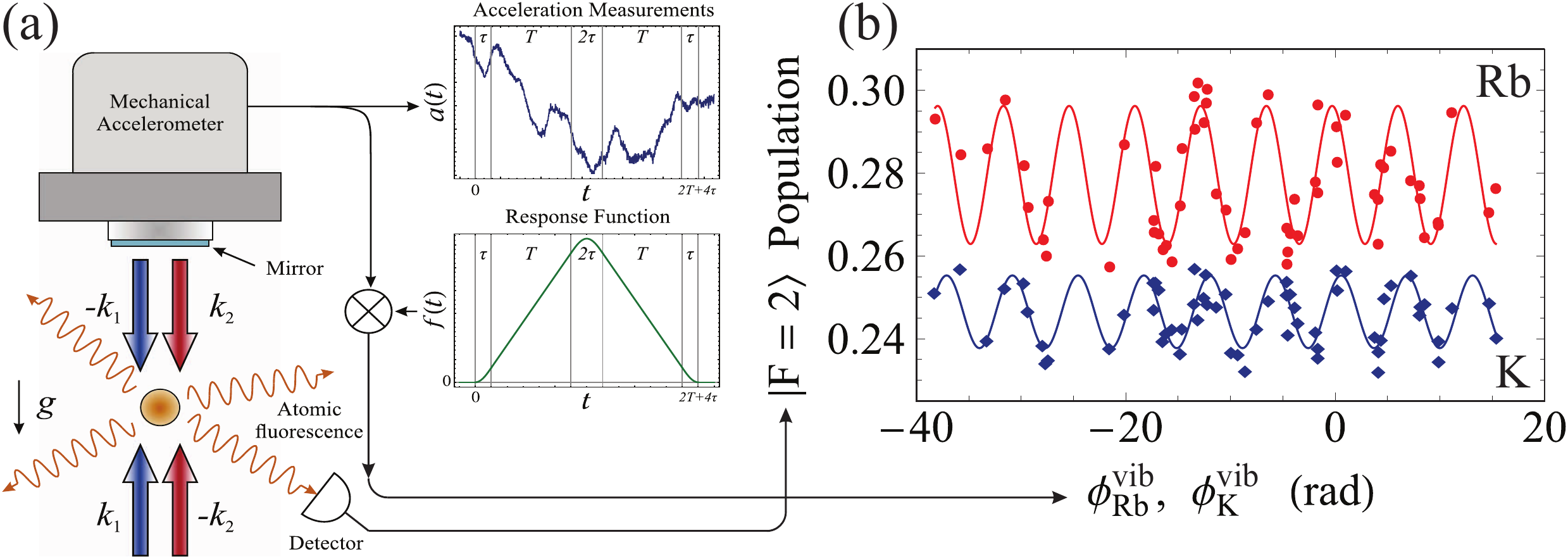}
  \caption{(a) Schematic of the FRAC method. The acceleration is recorded during the interferometer sequence and is convolved with the interferometer response function to obtain the vibration phase $\phi_{\mathrm{vib},j}$. (b) Simultaneous $^{87}$Rb (red) and $^{39}$K (blue) fringes obtained using the FRAC method in $0g$ with an interrogation time of $T = 2$ ms.}
  \label{fig:FRAC-0gFringes}
\end{figure*}

However, as one goes to larger sensitivities, or when the vibration noise is large (such as onboard a moving vehicle), the motion of the reference mirror becomes a significant source of phase noise. In most cases, one must employ methods to either suppress or reject the phase noise resulting from mirror vibrations. A common method to suppress ambient vibrations is to place the apparatus on an anti-vibration platform~\cite{LeGouet2008,Zhou2012}, but this solution is not generally compatible with onboard applications. Instead, we take two approaches to reject vibration noise. The first is to perform the interferometers simultaneously and with the same scale factor (\ie $\mathcal{S}_{\rm Rb} = \mathcal{S}_{\rm K}$), this ensures a maximum amount of common mode rejection. It is then possible to measure the differential phase from the ellipsoidal pattern obtained by parametrically plotting the interferometer output for each species~\cite{Barrett2015} [see \Fig \ref{fig:KRb-CorrelatedFringes}(b)]. The second method involves monitoring the mirror vibrations with a mechanical accelerometer and correcting for the noise by computing the corresponding vibration phase via~\cite{Geiger2011,Barrett2015}
\be
  \label{eq:vibsensitivity}
  \phi_{\mathrm{vib},j} = k_j \int_0^{2T_j} f_j(t) a_{\rm vib}(t) \mathrm{d}t,
\ee
where $f_j(t)$ is the interferometer response function and $a_{\rm vib}(t)$ is the common acceleration of the reference mirror due to vibrations during the interferometer sequence. We call this technique the fringe reconstruction by accelerometer correlation (FRAC) method [see \Fig \ref{fig:FRAC-0gFringes}(a)]. This latter approach is much more versatile when the mechanical accelerometer has sufficiently high signal-to-noise ratio in the desired frequency band, since it does not require the two scale factors to be equal, and it does not rely on complex ellipse fitting algorithms that can yield biased estimates of $\Delta\Phi_d$.

\subsection{Testing the WEP on ground}

Since the ICE apparatus is designed primarily to operate in microgravity, where the atoms remain close to the center of the science chamber, the maximum interrogation time we can achieve on ground is limited to $2T \simeq 45$ ms by our detection optics. Hence, to increase the sensitivity of our measurements, we have focused on improving the signal-to-noise ratio (SNR) of, and the degree of correlation between, both $^{87}$Rb and $^{39}$K interference fringes. A variety of recent improvements to our potassium source, including gray-molasses cooling and a new state preparation scheme, have enabled us to increase the fringe contrast and SNR by a factor of 4. Similarly, by using an architecture where the interrogation beams for each species travel along the same optical pathway, and have the same interaction strength with their respective atoms, we maximize the level of common-mode rejection and enhance sensitivity to differential effects.

Figure \ref{fig:KRb-CorrelatedFringes} shows simultaneous measurements from the laboratory using $T = 20$ ms. Here, ambient mirror vibrations are large enough to scan the interferometer fringes over a phase range of $\sim 4\pi$. Figure \ref{fig:KRb-CorrelatedFringes}(a) displays the raw transition probability of both species for 2000 accelerations measurements (requiring $\sim 30$ minutes of data acquisition). These points are integrated over the horizontal axis to obtain the probability density function. By fitting a function that is the convolution of an inverse sine and a Gaussian to the probability density, we extract the contrast and SNR of each interferometer~\cite{Geiger2011}. Figure \ref{fig:KRb-CorrelatedFringes}(b) displays a parametric plot of the potassium transition probability as a function of the rubidium one for the same data presented in \Fig \ref{fig:KRb-CorrelatedFringes}(a). Since the vibration noise is common mode for both species, there is a high degree of correlation between these data---resulting in an ellipse with an eccentricity determined by the differential phase between the two interference fringes. Fitting an ellipse to these data yields an estimate of $\Delta\Phi_d$ with a statistical precision of $\sim 10$ mrad---corresponding to a WEP violation sensitivity of $\delta\eta \sim 1.5 \times 10^{-7}$. Figure \ref{fig:KRb-CorrelatedFringes}(c) shows the reconstructed interference fringes using the FRAC method, from which we can directly measure the acceleration sensitivity of each species: $\delta a_{\rm Rb} \simeq 5.2 \times 10^{-8}\,g$ and $\delta a_{\rm K} \simeq 8.3 \times 10^{-8}\,g$, corresponding to a WEP sensitivity of $\delta\eta \simeq 7.2 \times 10^{-8}$ after 30 minutes of integration. Although, further study of systematic effects is required, these results are approaching precisions competitive with the state-of-the-art~\cite{prl5,Rosi2017}.

\subsection{Testing the WEP in weightlessness}

In May 2015, the ICE experiment embarked on a series of parabolic flight campaigns and successfully tested the WEP for the first time in weightlessness~\cite{Barrett2016}. Here, we give a review of our results and the issues that remain to be solved for future experiments.

When performing experiments in a moving vehicle, the primary challenge for an AI is to deal with the large level of vibration noise (in the case of an aircraft, this can be 5 orders of magnitude above that of a quiet laboratory), and changes in acceleration and orientation due to the vehicle's motion. Our approach to these problems has been to utilize a sensitive classical accelerometer to monitor and correct for high-frequency vibrations on the reference mirror via the aforementioned FRAC method [see \Fig \ref{fig:FRAC-0gFringes}(a)]. This sensor also gives us information about low-frequency changes in acceleration (particularly during the $0g$, $1g$ and $2g$ phases of parabolic flight) which enable us to appropriately adapt the phase modulation of our interrogation laser during the interferometer to optimize the fringe contrast~\cite{Barrett2016}. During these campaigns, we obtained correlated acceleration measurements from each species during microgravity [see \Fig \ref{fig:FRAC-0gFringes}(b)] that yielded a measurement of $\eta = (0.9 \pm 3.0) \times 10^{-4}$ at $T = 2$ ms. When compared to measurements made during $1g$ on the same flights, we find a 5-fold improvement as a result of the rejection of certain systematic effects in $0g$. Although with a modest precision, this WEP test is the first of its kind---namely with cold atoms free-falling in an Einstein elevator---and serves as a test bed for future experiments in Space~\cite{Aguilera2014}. During these flights we also demonstrated a record-low acceleration sensitivity of $3.4 \times 10^{-5}\,g$ per shot with our $^{87}$Rb interferometer operating at $T = 5$ ms---corresponding to a resolution 1600 times below the level of ambient acceleration noise. This opens the way toward inertial navigation based on cold-atom technology~\cite{Battelier2016}.

Presently, the sensitivity of atom-interferometric measurements on the aircraft is limited by it's rotational motion. When the rotation occurs along an axis perpendicular to that of the interrogation beams, the upper and lower pathways of the Mach-Zehnder interferometer begin to diverge and fail to overlap during the final recombination pulse. This effect results in a loss of contrast which scales as~\cite{Roura2014,Barrett2016}
\be
  \mathcal{C} \sim e^{-(k \sigma_v T)^2 (|\bm{\Omega}_{\rm tr}| T)^2},
\ee
where $\sigma_v$ is the width of the velocity distribution ($\sigma_v^2$ is proportional to the sample temperature) and $\bm{\Omega}_{\rm tr}$ is the rotation vector transverse to $\bm{k}$. In future experiments, we plan to counter-rotate the reference mirror in real-time to compensate for this effect.


\section{Conclusion}

Cold-atom interferometry is a very promising tool to reach the challenging sensitivities required for fundamental precision measurements such as the detection of gravitational waves or violations of the equivalence principle. Such techniques can be exploited to provide new measurement apparatuses from lab-sized to large-scale infrastructures. Beyond the impact of atom interferometry in fundamental physics, the development of such experiments brings new technologies opening applications in other fields such as geology, seismology and hydrology, for underground surveys and the detection of mass transfer, but also for the realization of quantum-sensor-based inertial navigation systems. In terms of the projects discussed here, The MIGA antenna will be a first prototype for a new type of GW detector in the 0.1--1 Hz frequency band, and the first large-scale infrastructure based on quantum technologies. We are now realizing a first-generation experiment based on an original cavity geometry that will allow the team to study future upgrades to the MIGA antenna. Similarly, the ICE project has achieved the first WEP tests in microgravity, and is presently pursuing a precision measurement of the E\"{o}tv\"{o}s parameter with a very-long-baseline interferometer. Toward this goal we are preparing an evaporative cooling stage to reach ultra-cold temperatures lower than $10$ nK. In addition, we have recently installed an Einstein elevator in the lab which gives access to $\sim 500$ ms of weightlessness every 10 s. With these new tools, we anticipate a measurement of $\eta$ below $10^{-10}$ to be within reach.


\section*{Acknowledgments}

This work was realized with the financial support of the French State through the ``Agence Nationale de la Recherche" (ANR) in the frame of the ``Investissement d'avenir" programs: Equipex MIGA (ANR-11-EQPX-0028), IdEx Bordeaux - LAPHIA (ANR-10-IDEX-03-02) and FIRST-TF (ANR-10-LABX-48-01). This work is also supported by the French national agencies CNES, DGA, IFRAF, action sp\'ecifique GRAM, RTRA ‘Triangle de la Physique’ and the European Space Agency. This work was also supported by the R\'egion d'Aquitaine (project IASIG-3D) and by the city of Paris (Emergence project HSENS-MWGRAV).  We also thank the ``P\^ole de comp\'etitivit\'e Route des lasers - Bordeaux'' cluster for his support. G. L. thanks DGA for financial support. M. P. also thanks LAPHIA--IdEx Bordeaux for partial financial support.


\end{document}